\documentclass[11pt]{article}
\usepackage{latexsym}
\usepackage{amsmath,amsfonts,amsbsy,amssymb,amscd}
\usepackage{fancybox,fancyhdr}
\usepackage{cite}
\usepackage{pstricks,pst-node,psfrag}
\usepackage{graphicx,epsfig,picture,graphics}
\usepackage{slashed}
\usepackage[utf8]{inputenc}

\psset{unit=1.3cm,linewidth=.5pt,radius=.2}  

\usepackage{float}                          
\usepackage{lscape}                         
\usepackage{bm}
\usepackage{tikz}

\newcommand{\hoch}[1]{$\, ^{#1}$}

\newsavebox{\uuunit}
\sbox{\uuunit}
    {\setlength{\unitlength}{0.825em}
     \begin{picture}(0.6,0.7)
        \thinlines
        \put(0,0){\line(1,0){0.5}}
        \put(0.15,0){\line(0,1){0.7}}
        \put(0.35,0){\line(0,1){0.8}}
       \multiput(0.3,0.8)(-0.04,-0.02){12}{\rule{0.5pt}{0.5pt}}
     \end {picture}}
\newcommand {\unity}{\mathord{\!\usebox{\uuunit}}}

\usepackage{color}
\definecolor{MyDarkBlue}{rgb}{0.15,0.15,0.45}    \definecolor{MyGreen}{rgb}{0.15,0.45,0.45}
\definecolor{MyPurple}{rgb}{0.55,0.25,0.55}
\usepackage[linktocpage=true]{hyperref}
\hypersetup{colorlinks=true,citecolor=MyGreen,linkcolor=MyPurple,urlcolor=MyGreen}

\begin{document}

\begin{flushright}
\hfill{ UG-15-84}
\end{flushright}
\vskip 2.5cm

\begin{center}
{\Large \bf Non-relativistic fields from arbitrary contracting backgrounds}
\end{center}
\vspace{27pt}
\begin{center}
{\Large {\bf }}

\vspace{15pt}

{\Large Eric Bergshoeff\hoch{1}, Jan Rosseel\hoch{2} and Thomas Zojer\hoch{1}}

\vspace{32pt}

\hoch{1} {\it Van Swinderen Institute for Particle Physics and Gravity, University of Groningen,\\
Nijenborgh 4, 9747 AG Groningen, The Netherlands}\\

\vspace{5pt}

\texttt{e.a.bergshoeff@rug.nl, t.zojer@rug.nl} \\

\vspace{10pt}

\hoch{2} {\it Albert Einstein Center for Fundamental Physics, University of Bern, \\
Sidlerstrasse 5, 3012 Bern, Switzerland}\\

\vspace{5pt}

\texttt{rosseel@itp.unibe.ch}

\vspace{65pt}
\underline{ABSTRACT}
\end{center}

\bigskip
\noindent
We discuss a non-relativistic contraction of massive and massless field theories minimally coupled to gravity. Using the non-relativistic limiting procedure introduced in our previous work, we (re-)derive non-relativistic field theories of massive and massless spins 0 to 3/2 coupled to torsionless Newton-Cartan backgrounds. We elucidate the relativistic origin of the Newton-Cartan central charge gauge field $m_\mu$ and explain its relation to particle number conservation.

\vspace{15pt}

\thispagestyle{empty}

\vspace{15pt}

 \vfill

\voffset=-40pt

\newpage

\tableofcontents




\section{Introduction \label{sec:intro}}

Recent years have witnessed an increased appreciation for formulations
of non-relativistic field theories in arbitrary space-time backgrounds
and coordinate frames. Coupling non-relativistic field theories to
arbitrary backgrounds has practical benefits, such as enabling one to
easily define conserved currents like the energy-momentum tensor and
to study linear response. Furthermore, diffeomorphism invariance and
background independence are also useful tools in studying strongly
coupled condensed matter systems, for which a perturbative expansion
in powers of a small coupling constant is not at hand. One then often
resorts to an effective field theory approach or to holographic
methods. In the former case, one writes hydrodynamical equations or
effective Lagrangians in an expansion that is organized in number of
derivatives. Requiring invariance under general coordinate
transformations and local space-time symmetries that are present in
arbitrary backgrounds is then of great use in restricting the number
of possible terms. Such effective field theory methods have been used
in the description of e.g.~the unitary Fermi gas \cite{Son:2005rv} or
the quantum Hall effect
\cite{Hoyos:2011ez,Son:2013rqa,Abanov:2013woa,Golkar:2014wwa,Geracie:2014nka}. A
holographic approach on the other hand, rephrases the calculation of
e.g.~partition functions as a boundary value problem for a dual
gravitational theory in a space-time with non-relativistic boundary
and isometry group
\cite{Son:2008ye,Kachru:2008yh,Balasubramanian:2008dm,Hartnoll:2009sz}. Introducing
arbitrary non-relativistic geometric boundary data is then important
as these are interpreted as sources for e.g.~the energy-momentum
tensor of the strongly coupled field theory under study (see
e.g.~\cite{Christensen:2013lma,Christensen:2013rfa,Hartong:2014oma,Hartong:2015wxa,Hartong:2014pma}
for a discussion in the context of Lifshitz holography).

In the above mentioned applications and approaches, it is important to
understand how a field theory can be consistently coupled to an
arbitrary non-relativistic background. This is the problem we will be
concerned with in this paper. For relativistic field theories, a
well-known way of introducing background independence and
diffeomorphism invariance is given by using minimal coupling,
i.e.~replacing the Minkowski metric by an arbitrary Lorentzian
metric. Here, we will be concerned with a similar procedure of minimal
coupling for non-relativistic theories and we will show that this can
be obtained as a limit from a particular relativistic minimal coupling
procedure.

Before doing this, one has to address what the
non-relativistic analog of Lorentzian geometry is to which one wishes
to couple the field theory. As has been argued in both the condensed
matter and applied holography literature, the correct geometric
framework is the one offered by Newton-Cartan geometry, with or
without torsion. Newton-Cartan geometry refers to a
differential-geometric framework for Newtonian space-times that can be
used to recast Newtonian gravity in a manner that resembles General
Relativity. Originally devised in a metric-like formulation,
Newton-Cartan geometry in the absence and presence of torsion has
recently been rewritten using vielbeins
\cite{Andringa:2010it,Andringa:2013mma,Bergshoeff:2014uea}
by gauging the Bargmann algebra, i.e.~the central extension of the
Galilei algebra of non-relativistic space-time symmetries. Apart from
a time-like vielbein $\tau_\mu$
and a space-like vielbein $e_\mu{}^a$,
one also needs to include a one-form $m_\mu$.
The latter can be interpreted as a gauge field for the central charge
of the Bargmann algebra that is related to particle number
conservation.  As in the relativistic case, a vielbein formulation of
the background geometry is essential to consider for instance
couplings to fermion fields, due to the manifest presence of local
spatial rotations and Galilean boosts. For this reason, it will also
be our preferred choice for studying minimally coupled
non-relativistic field theories.

In \cite{Bergshoeff:2015uaa}, it was shown how the vielbein
formulation of torsionless Newton-Cartan geometry arises from deforming an
arbitrary relativistic background to a Newton-Cartan one using a
procedure that mimics the In\"on\"u-Wigner contraction of the
Poincar\'e algebra to the Bargmann algebra. In this ``background
contraction'', one starts from a relativistic background, that is not
only given by the gravitational field (the vielbein $E_\mu{}^A$),
but that also includes a U(1) gauge field $M_\mu$.
The inclusion of the latter is motivated by the fact that the
contraction from $\{E_\mu{}^A, M_\mu\}$
to the Newton-Cartan background $\{\tau_\mu, e_\mu{}^a, m_\mu\}$
then conveniently preserves the number of field components. In a first
step, the contraction consists of redefining the fields $E_\mu{}^A$,
$M_\mu$
in terms of the eventual Newton-Cartan ones $\tau_\mu$,
$e_\mu{}^a$,
$m_\mu$
and a contraction parameter $\omega$.
This contraction parameter is then taken to infinity in a second
step. In order for this limit to make sense, the gauge field $M_\mu$
is constrained to be a flat connection, guaranteeing that all
necessary background geometric quantities, such as the spin
connection, have a well-defined $\omega \to \infty$ limit.
Using this background contraction, one can then derive the
kinematics of Newton-Cartan gravity from that of General Relativity and similar results for three-dimensional Newton-Cartan supergravity, as was shown in \cite{Bergshoeff:2015uaa}.

In this paper, we will argue that this procedure can be easily
extended to obtain non-relativistic field theories that are minimally
coupled to arbitrary torsionless Newton-Cartan backgrounds from
minimally coupled relativistic field theories. We will do this
starting from Lagrangians for relativistic theories that describe
massive and massless free fields with spins ranging from 0 up to
3/2. We will then show that the $\omega$-dependent
redefinitions of the background fields mentioned above, can be
supplemented with similar redefinitions of the dynamical fields in
such a way that the Lagrangians have a well-defined
$\omega \to \infty$ limit
after performing all field redefinitions. Applying this limit then
leads to non-relativistic theories minimally coupled to arbitrary
torsionless Newton-Cartan backgrounds.

The background gauge field $M_\mu$
will play a particular role in this paper when discussing massive
fields. Indeed, since massive non-relativistic fields typically obey a
Schr\"odinger-like equation, they are complex. In order to obtain such
theories, we will also start from Lagrangians for complex relativistic
fields, that thus feature a U(1) invariance, whose associated Noether
current expresses conservation of the number of particles minus the
number of antiparticles. We will then include two types of minimal
couplings in these Lagrangians. On the one hand, we will introduce the
usual gravitational minimal couplings, by replacing the Minkowski
metric by an arbitrary Lorentzian one. On the other hand, we will also
gauge the U(1) symmetry, by introducing minimal couplings to the
background gauge field $M_\mu$ \footnote{A similar procedure of using minimal couplings to a flat background gauge field to obtain a well-defined non-relativistic limit has already appeared in the context of non-relativistic particle and string actions in flat space-times in \cite{Gomis:2000bd}.}.
We will then show that the latter couplings make it possible to suitably extend the background contraction to
theories with dynamical massive fields. 
For massless fields, we will see that there is no need to require a U(1) symmetry and associated minimal couplings to $M_\mu$ in order to be able to extend the background
contraction. One can thus start from relativistic Lagrangians for real massless fields,
that only contain gravitational minimal couplings.

Other approaches to define the non-relativistic limit of arbitrary relativistic backgrounds have been considered in the literature. Some work on Newton-Cartan geometry in the metric formulation as the
non-relativistic limit of General Relativity can be found in
\cite{Dautcourt,Kuenzle,Kuchar:1980tw,Bleeken:2015ykr}. Regarding field theories
coupled to Newton-Cartan geometry, most of the literature so far has
focused on massive spin-0 and spin-1/2 fields and here we also
provide results for fields of spin-1 and spin-3/2 and for massless
fields. Of the recent literature, we mention in particular the work
\cite{Jensen:2014wha,Fuini:2015yva,Geracie:2015dea}. The present work
differs from \cite{Geracie:2015dea} in the introduction of the flat
connection $M_\mu$
in the relativistic starting point. A similar relativistic background
gauge field is introduced in the work of
\cite{Jensen:2014wha,Fuini:2015yva}. The limit discussed here differs
however from \cite{Jensen:2014wha,Fuini:2015yva} in the redefinitions
we perform on the background vielbein and $M_\mu$
before taking the $\omega \to \infty$
limit. Further differences stem from the fact that here we restrict
ourselves to torsionless Newton-Cartan geometry.

The outline of this paper is as follows. In section \ref{sec:bgcontr} we review the background contraction of \cite{Bergshoeff:2015uaa}. In section \ref{sec:massive} we show how this procedure can be extended to obtain non-relativistic field theories in arbitrary torsionless Newton-Cartan backgrounds from suitably minimally coupled relativistic theories. The case of massless fields is explained in section \ref{sec:massless}. We end with conclusions and an outlook on further research in section \ref{sec:concl}.

\section{Contracting relativistic backgrounds} \label{sec:bgcontr}

In \cite{Bergshoeff:2015uaa}, it was explained how a torsionless Newton-Cartan background can be obtained from a relativistic one, via a procedure that mimics the In\"on\"u-Wigner contraction from the Poincar\'e to the Bargmann algebra. This ``background contraction'' can thus be viewed as the non-relativistic limit of a relativistic background, in the same way as the Bargmann algebra can be regarded as the non-relativistic limit of the Poincar\'e algebra. Here, we will give a brief review of this procedure, emphasizing the points that are relevant for this paper. We refer to \cite{Bergshoeff:2015uaa} for further details, as well as applications to Newton-Cartan (super)gravity. Here and in the following, we will use uppercase Latin indices $A$, $B$ to denote relativistic flat indices. When making a time-space split, we will reserve $0$ for the time-like index and we will use lowercase Latin indices $a$, $b$ for the spatial directions.

The relativistic background one starts from is specified by two fields: the vielbein $E_\mu{}^A$, that represents the gravitational background, as well as a U(1) gauge field $M_\mu$. As explained in \cite{Bergshoeff:2015uaa}, the inclusion of the latter is motivated by the In\"on\"u-Wigner contraction used to derive the Bargmann algebra. This contraction is most conveniently performed on the direct sum of the Poincar\'e algebra with an extra abelian generator and mimicking this leads one to also include an extra U(1) gauge field in the background. Importantly, the field $M_\mu$ is constrained to be curl-free, i.e.~a flat connection:
\begin{equation} \label{eq:Mnocurl}
  \partial_{[\mu} M_{\nu]} = 0 \,.
\end{equation}
Imposing this constraint ensures that no extra degrees of freedom are introduced in the background, apart from the gravitational ones. Furthermore, this constraint also ensures that background geometric quantities are well-defined under the contraction procedure.

Starting from $E_\mu{}^A$ and $M_\mu$ the background contraction is then performed as follows. One first introduces fields $\tau_\mu$, $e_\mu{}^a$ and $m_\mu$ via the following field redefinitions, that involve a contraction parameter $\omega$:
\begin{align}
  \label{eq:redefEM}
  E_\mu{}^0 &= \omega\, \tau_\mu + \frac{1}{2\omega}\, m_\mu \,, \qquad \qquad M_\mu = \omega\, \tau_\mu - \frac{1}{2\omega}\, m_\mu \,, \qquad \qquad E_\mu{}^a = e_\mu{}^a \,.
\end{align}
The fields $\tau_\mu$, $e_\mu{}^a$ and $m_\mu$ correspond to the time-like vielbein, spatial vielbein and central charge gauge field of the Newton-Cartan background that will result from the contraction.
The inverse vielbein $E^\mu{}_A$ can then be expressed as an infinite series in $\omega$. Up to the order that is needed, this expansion is given by
\begin{align}
  \label{eq:redefinvE}
  E^\mu{}_0 = \frac{1}{\omega}\, \tau^\mu - \frac{1}{2\omega^3}\, \tau^\mu \tau^\rho m_\rho + \mathcal{O}\left(\omega^{-5}\right) \,, \qquad
  E^\mu{}_a = e^\mu{}_a - \frac{1}{2 \omega^2}\, \tau^\mu e^\rho{}_a m_\rho + \mathcal{O}\left(\omega^{-4}\right) \,,
\end{align}
where we have introduced fields $\tau^\mu$, $e^\mu{}_a$, that will correspond to inverse Newton-Cartan vielbeine, obeying
\begin{align}\begin{aligned}
  \label{eq:invNCvielbrels}
   \tau^\mu \tau_\mu &= 1 \,, & \tau^\mu e_\mu{}^a &= 0 \,, \qquad \qquad & e^\mu{}_a \tau_\mu &= 0 \,, \\
   e_\mu{}^a e^\nu{}_a &= \delta^\nu_\mu - \tau^\nu \tau_\mu \,, \qquad \qquad & e^\mu{}_a e_\mu{}^b &= \delta^b_a \,.
\end{aligned}\end{align}
Applying the redefinitions (\ref{eq:redefEM}) to the determinant $E$ of the vielbein $E_\mu{}^A$ leads to
\begin{align}
  \label{eq:redefdetE}
  E = {\rm det}(E_\mu{}^A) = \omega\, {\rm det}\left(\tau_\mu, e_\mu{}^a\right) + \frac{1}{2\omega}\, {\rm det}\left(m_\mu, e_\mu{}^a\right) \,.
\end{align}
In the following, only the term involving ${\rm det}\left(\tau_\mu, e_\mu{}^a\right)$ will be relevant. With slight abuse of notation, we will denote this determinant by $e$:
\begin{align}
  \label{eq:defe}
  e \equiv {\rm det}\left(\tau_\mu, e_\mu{}^a\right) \,.
\end{align}
Note that $e$ represents an appropriate volume element that transforms as a density under general coordinate transformations (and is invariant under the local Galilean boosts, spatial rotations and central charge transformations of (\ref{eq:NCsymmrules})). We also note that one can introduce an affine connection and appropriate vielbein postulates for $\tau_\mu$ and $e_\mu{}^a$ \cite{Andringa:2010it}. Using these, one can show that the volume element (\ref{eq:defe}) allows one to partially integrate similarly as in Einstein-Hilbert gravity.

The redefinitions (\ref{eq:redefEM}), consequences (\ref{eq:redefinvE}) and the constraint (\ref{eq:Mnocurl}) can be used in the usual expression for the relativistic spin connection $\Omega_\mu{}^{AB}(E)$ in terms of the vielbein $E_\mu{}^A$, to obtain an expansion in $\omega$
\begin{align}
  \label{eq:spinconnexp}
  \Omega_\mu{}^{ab}(E) &= \omega_\mu{}^{ab}(e,\tau,m) + \mathcal{O}\left(\omega^{-2}\right) \,, \qquad
  \Omega_\mu{}^{0a}(E) = \frac{1}{\omega}\, \omega_\mu{}^a (e,\tau,m) + \mathcal{O}\left(\omega^{-3}\right) \,.
\end{align}
The expressions $\omega_\mu{}^{ab}(e,\tau,m)$, $\omega_\mu{}^a(e,\tau,m)$ that appear in the leading order terms coincide with those for the dependent spin connections of spatial rotations and Galilean boosts in torsionless Newton-Cartan geometry \cite{Andringa:2010it,Andringa:2013mma,Bergshoeff:2015uaa}.

The contraction procedure on a quantity containing relativistic background geometry fields is then effectuated by expressing everything in terms of $\tau_\mu$, $e_\mu{}^a$, $m_\mu$, using the above redefinitions and their consequences and taking the limit $\omega \rightarrow \infty$. 
The contraction can also be applied to the symmetries under which $E_\mu{}^A$ and $M_\mu$ transform. These are given by diffeomorphisms, local Lorentz rotations and the U(1) transformation of $M_\mu$. The contraction leaves one with diffeomorphisms (with parameter $\xi^\mu$), local spatial rotations (with parameter $\lambda^{ab}$), local Galilean boosts (with parameter $\lambda^a$) and a local central charge transformation (with parameter $\sigma$). Under these symmetries, the Newton-Cartan fields obey the following transformation rules
\begin{align}
  \label{eq:NCsymmrules}
  \delta \tau_\mu &= \xi^\nu \partial_\nu \tau_\mu + \partial_\mu \xi^\nu \tau_\nu \,, \nonumber \\
  \delta e_\mu{}^a &= \xi^\nu \partial_\nu e_\mu{}^a + \partial_\mu \xi^\nu e_\nu{}^a + \lambda^a{}_b e_\mu{}^b + \lambda^a \tau_\mu \,, \nonumber \\
  \delta m_\mu &= \xi^\nu \partial_\nu m_\mu + \partial_\mu \xi^\nu m_\nu + \partial_\mu \sigma + \lambda^a e_{\mu\, a} \,.
\end{align}
It will in the following sometimes be interesting to restrict an arbitrary background to a flat one, defined by the following values for the Newton-Cartan background fields
\begin{align}
  \label{eq:flatbackground}
  \tau_\mu = \delta_\mu^t \,, \qquad\quad e_t{}^a = 0,\quad e_i{}^a = \delta_i^a \,, \qquad\quad m_\mu = 0 \,.
\end{align}
The subset of the symmetries (\ref{eq:NCsymmrules}) that leaves these flat background values invariant can be found by solving the following non-relativistic Killing equations
\begin{align}\begin{aligned}
  \label{eq:killeqsnr}
   \partial_\mu \xi^t &= 0 \,, \qquad &\partial_t \xi^i + \lambda^i &= 0 \,, \\
   \partial_i \xi^j + \lambda^j{}_i &= 0 \,, \qquad& \partial_t \sigma &= 0 \,, \qquad& \partial_i \sigma + \lambda_i &= 0 \,.
\end{aligned}\end{align}
The solution to these equations is given by
\begin{align}
  \label{eq:killvecsnr}
  \xi^t (x^\mu) = \zeta\,, \qquad \xi^i(x^\mu) = \xi^i - \lambda^i\, t - \lambda^i{}_j\, x^j\,, \qquad \sigma(x^\mu) = \sigma - \lambda^i \, x^i\,,
\end{align}
where the parameters $\zeta$, $\xi^i$, $\lambda^i$, $\lambda^{ij}$, $\sigma$ are now constants. These correspond to the usual time translation, spatial translations, Galilean boosts, spatial rotations and central charge transformation of the rigid Bargmann algebra.

Using this background contraction, it was shown in \cite{Bergshoeff:2015uaa} that the kinematics of Newton-Cartan (super)gravity can be derived from the kinematics of relativistic (super)gravity.
In the following two sections, we will extend this result to Lagrangians for massive and massless dynamical fields of spin-0 up to spin-3/2 in arbitrary backgrounds. We will in particular show that starting from a relativistic Lagrangian that includes appropriate minimal couplings to $E_\mu{}^A$ and $M_\mu$, the above redefinitions (\ref{eq:redefEM}) can be extended with similar redefinitions for the dynamical fields in such a way that the $\omega \rightarrow \infty$ limit is well-defined and leads to Lagrangians that are coupled to an arbitrary Newton-Cartan background.

\section{Massive fields} \label{sec:massive}

In this section, we will discuss the coupling of massive fields with
spins 0 to 3/2 to an arbitrary torsionless Newton-Cartan background
and how these results are obtained from minimally coupled massive
relativistic field theories.  As discussed in
\cite{Bargmann:1954gh,Leblond:1963,LevyLeblond:1967zz}, non-trivial massive
representations of the Galilei group are not true but rather ray
representations. They correspond to true representations of the
Bargmann algebra with non-trivial central charge transformations and
are thus associated to complex fields, that acquire non-trivial phase
factors under the action of the central charge. When considered in a
flat background such fields typically obey Schr\"odinger-like
equations. The ray character of the representation or the effect of
the central charge can then be seen from the fact that the field
transforms with a non-trivial phase factor under rigid Galilean
boosts.

In an arbitrary curved background, the rigid space-time
transformations of the Galilei group are replaced by local
diffeomorphisms, local spatial rotations and local Galilean
boosts. The fact that massive fields correspond to true
representations of the Bargmann algebra instead of the Galilei
algebra suggests that the central charge phase transformations also
become local. Indeed, the description of Newton-Cartan geometry
contains the field $m_\mu$
whose role is precisely to gauge the central charge. A consistent
coupling of a field to an arbitrary Newton-Cartan background can
therefore be expected to include couplings to $m_\mu$,
apart from couplings to the background vielbeine $\tau_\mu$,
$e_\mu{}^a$.

In order to obtain such central charge gauge couplings, we will start from relativistic Lagrangians for
complex fields. These are invariant under a U(1) symmetry whose
associated Noether current describes conservation of the number of
particles minus the number of antiparticles. We will then introduce
minimal couplings of two different types. On the one hand, we
minimally couple the Lagrangians to an arbitrary gravitational
background by replacing the Minkowski metric by a generic Lorentzian
one. On the other hand, we will also minimally couple them to the
background gauge field $M_\mu$
thus gauging the U(1) symmetry. Introducing the latter couplings will
enable us to apply the background contraction (suitably extended with
$\omega$-dependent
redefinitions of the dynamical fields) to these systems in a
well-defined way. Indeed, we will see that the mass terms typically
diverge in the $\omega \to \infty$
limit and it turns out that these divergences can be cancelled against
similar divergences stemming from the couplings to $M_\mu$.
Moreover, the minimal couplings to $M_\mu$
will also lead to the required couplings to $m_\mu$
in the resulting non-relativistic theories. Since the central charge
phase transformation, that is gauged by $m_\mu$
is related to particle number conservation, one sees that the
background contraction has the effect of suppressing either
antiparticles or particles.

In the following, we will illustrate this procedure for massive free
fields of spin-0 up to spin-3/2. In each case, we will give the
resulting non-relativistic theories in an arbitrary background and
for the bosonic fields we discuss the restriction to flat backgrounds
as well.

\subsection{Spin-0}

As explained above, we start from a Lagrangian for a relativistic complex scalar $\Phi$, minimally coupled to an arbitrary gravitational background and the extra U(1) gauge field $M_\mu$:
\begin{align}\label{relscalar}
 E^{-1} \,\mathcal{L}_{\rm rel} = -\frac12\,g^{\mu\nu}\,D_\mu\Phi^*D_\nu\Phi -\frac{M^2}{2}\,|\Phi|^2 \,,
\end{align}
where the covariant derivative is given by
\begin{align}
 D_\mu\Phi = \partial_\mu\Phi - {\rm i}\,M\,M_\mu\,\Phi \,.
\end{align}
Apart from invariance under diffeomorphisms, the above Lagrangian is thus also invariant under a local U(1) symmetry given by the transformation rule
\begin{equation}
  \label{eq:PhiU1}
  \delta \Phi = {\rm{i}}\, M\, \Lambda\, \Phi \,.
\end{equation}
Note that the current associated to this local U(1) symmetry is given by
\begin{equation}
  j^\mu_{\rm rel} = E^{-1}\frac{\delta \mathcal{L}_{\rm rel}}{\delta M_\mu} = \frac{M}{2{\rm{i}}} \, \Big( \Phi^* D^\mu \Phi - \Phi D^\mu \Phi^*\Big) \,.
\end{equation}
This current is thus proportional to the usual current that expresses conservation of the number of particles minus the number of antiparticles. Often, this current is identified with an electromagnetic one. Here however, we do not wish to make this identification since we can not assume that $M_\mu$ is an electromagnetic field. Indeed, as stressed above, in order to have a well-defined limit for the spin connection of the background geometry, we have to impose that $M_\mu$ is curl-free and hence a flat connection. It is possible to introduce a coupling to electromagnetism, but then an additional electromagnetic gauge field $A_\mu$ and minimal coupling to $\Phi$ has to be introduced in (\ref{relscalar}). We will not do this here. We will however consider an example of how this is done in the next subsection, discussing the spin-1/2 case.

Using the redefinitions and expansions of section \ref{sec:bgcontr} in (\ref{relscalar}) and rescaling the mass parameter $M$ as
\begin{equation}
  \label{eq:massresc}
  M = \omega m \,,
\end{equation}
one can check that the $\omega \rightarrow \infty$ limit is well-defined and leads to the following Lagrangian\footnote{We have ignored an overall factor of $\omega$ coming from the redefinition of $E = \omega \, e + \mathcal{O}(\omega^{-1})$. This factor is irrelevant as it amounts to an overall rescaling of the Lagrangian and can be absorbed in a simple rescaling of the field with a factor of $\sqrt{\omega}$. Here, and in the following, we will for simplicity ignore this factor.} \footnote{Here, and in the following, we have turned curved indices into flat indices using the inverse Newton-Cartan vielbeine. Thus $\tilde{D}_0$, $\tilde{D}_a$ are shorthand for $\tau^\mu \tilde{D}_\mu$, $e^\mu{}_a \tilde{D}_\mu$. Spatial flat indices are raised and lowered with a Kronecker delta.}
\begin{align}\begin{split}\label{nrscalar}
e^{-1} \mathcal{L}_{\rm non-rel} = m\,\Big[\,\frac{{\rm{i}}}2\,\Big(\Phi^*\tilde D_0\Phi -\Phi\tilde D_0\Phi^*\Big)
             -\frac1{2m}\,\big|\tilde D_a\Phi\big|^2 \,\Big]  \,,
\end{split}\end{align}
where we have used
\begin{align}
 \tilde D_\mu\Phi = \partial_\mu\Phi +{\rm{i}}\,m\,m_\mu\,\Phi \,.
\end{align}
This Lagrangian is easily recognized as the Lagrangian for a Schr\"odinger field coupled to an arbitrary Newton-Cartan background. As such, it is invariant under diffeomorphisms (with parameter $\xi^\mu$) and the local U(1) central charge transformation of the Bargmann algebra (with parameter $\sigma$), under which $\Phi$ transforms as
\begin{equation} \label{eq:nonrelU1Phi}
  \delta \Phi = \xi^\mu \partial_\mu \Phi  -{\rm{i}}\, m \, \sigma \, \Phi \,.
\end{equation}
One can then define the current associated to the central charge transformation by
\begin{equation}
  \label{eq:partconscurrent}
  j^\mu_{\rm non-rel} = -\frac{e^{-1}}{m^2} \frac{\delta \mathcal{L}_{\rm non-rel}}{\delta m_\mu} =  \tau^\mu \,|\Phi|^2 + e^\mu{}_a\, \frac{1}{2 m {\rm{i}}} \left( \Phi^*\tilde{D}^a \Phi - \Phi\tilde{D}^a \Phi^*  \right)\,.
\end{equation}
When choosing a flat background, this corresponds to the usual current of particle number or mass conservation. So, even though in our relativistic starting point we gauged the symmetry associated to conservation of the number of particles minus the number of antiparticles, non-relativistically we end up with only particle number conservation. We thus explicitly see that, as expected for a non-relativistic limit, our procedure has suppressed either particles or antiparticles.

It is also instructive to look at the action on $\Phi$ of the symmetries that are left when a flat background is chosen. As explained in the previous section, the transformation rules (\ref{eq:nonrelU1Phi}) then reduce to
\begin{align} \label{eq:flatsymmscalnr}
 \delta\Phi = \Big(\zeta\,\partial_t +\xi^i\partial_i -\lambda^i\,t\,\partial_i -x^j\lambda^i{}_j\,\partial_i
              -{\rm{i}}\,m\,\sigma + {\rm{i}}\,m\,\lambda^ix^i\Big)\Phi \,.
\end{align}
In particular, the last term in this transformation rule corresponds to the phase factor acquired by a Schr\"odinger field under rigid Galilean boosts, that is necessary to show Galilei invariance of the flat space Schr\"odinger equation that is obeyed by $\Phi$. Note that this Schr\"odinger equation is also invariant under an extra dilatation and special conformal transformation, that extend the symmetries of the Bargmann algebra denoted in (\ref{eq:flatsymmscalnr}) to the ones of the Schr\"odinger algebra. So, even though we started from a relativistic theory with no conformal invariance, the resulting non-relativistic theory is symmetric under Schr\"odinger symmetries.

\subsection{Spin-1/2}

Here, and also in other fermionic cases described in this paper, we will for convenience work in four space-time dimensions. The non-relativistic spin-1/2 case can then be discussed starting from the Lagrangian for a massive relativistic spinor $\Psi$ minimally coupled to the U(1) gauge field $M_\mu$ and an arbitrary gravitational background:
\begin{align}\label{relspin}
E^{-1} \mathcal{L}_{\rm rel} = \bar\Psi\slashed{D}\Psi -M\,\bar\Psi\Psi +{\rm h.c.}\,,
\end{align}
where  the covariant derivative is given by
\begin{align}
 D_\mu\Psi = \partial_\mu\Psi - \frac14\,\Omega_\mu{}^{AB}\gamma_{AB}\Psi + {\rm{i}}\,M\,M_\mu\,\Psi \,.
\end{align}
When we take the limit it is convenient to work with projected spinors $\Psi_{\pm}$ that transform nicely under the Galilean boosts as given in eq.~\eqref{nice}.
To be precise, we define these projected spinors  in terms of the original spinor as follows:
\begin{align} \label{eq:spinproj}
 \Psi_\pm  = \frac12\,\Big(\unity \pm {\rm{i}}\,\gamma^0\Big)\Psi\,, \hskip2cm \bar\Psi_\pm = \bar\Psi \,\frac12\,\Big(\unity \pm {\rm{i}}\,\gamma^0\Big) \,.
\end{align}
The background contraction of section \ref{sec:bgcontr} is then extended by redefining these projected spinors as follows:
\begin{align} \label{eq:spinprojresc}
 \Psi_+ = \sqrt{\omega}\,\psi_+ \,, \hskip2cm \Psi_- = \frac1{\sqrt\omega}\,\psi_- \,.
\end{align}
Using these redefinitions, along with the ones of section \ref{sec:bgcontr} and $M = \omega m$, one finds that the action \eqref{relspin} upon taking the $\omega\rightarrow\infty$ limit reduces to
\begin{align}\begin{split}\label{4dNRdirac}
e^{-1} \mathcal{L}_{\rm non-rel} &= \bar\psi_+\gamma^0\tilde D_0\psi_+ +\bar\psi_+\gamma^a\tilde  D_a\psi_- +\bar\psi_-\gamma^a\tilde D_a\psi_+
       -2\,m\,\bar\psi_-\psi_- +{\rm h.c.}\,,
\end{split}\end{align}
where we have used the covariant derivatives
\begin{align}\begin{split}\label{4dcovD}
 \tilde D_\mu\psi_+ &= \partial_\mu\psi_+ -\frac14\,\omega_\mu{}^{ab}\gamma_{ab}\psi_+ -{\rm{i}}\,m\,m_\mu\,\psi_+ \,, \\
 \tilde D_\mu\psi_- &= \partial_\mu\psi_- -\frac14\,\omega_\mu{}^{ab}\gamma_{ab}\psi_- +\frac12\,\omega_\mu{}^a\gamma_{a0}\psi_+
                     -{\rm{i}}\,m\,m_\mu\,\psi_- \,.
\end{split}\end{align}
The  invariance of the Lagrangian \eqref{4dNRdirac} under Galilean boosts is not manifest but can be checked by using the transformation rules
\begin{align}\begin{split}\label{nice}
 \delta \psi_+ &= \frac14\,\lambda^{ab}\gamma_{ab}\psi_+ + {\rm{i}} \, m \, \sigma \psi_+ \,, \\
 \delta \psi_- &= \frac14\,\lambda^{ab}\gamma_{ab}\psi_- -\frac12\,\lambda^a\gamma_{a0}\psi_+ + {\rm{i}} \, m \, \sigma \,\psi_- \,,
\end{split}\end{align}
that are easily found by applying all field redefinitions and taking the limit $\omega \to \infty$.

As mentioned above, the gauge field $M_\mu$
is not an electromagnetic one. It is however useful to consider a spinor that is charged under an additional
electromagnetic U(1). This extra U(1), with corresponding
gauge field ${A}_\mu$
and charge $q$,
should be distinguished from the central charge U(1), with gauge field
$M_\mu$
and charge $M$ (the mass parameter of the relativistic spinor).
The above contraction procedure can be extended to include this relativistic electromagnetic coupling, where there is no need to redefine $A_\mu$ with an $\omega$-dependent scale factor. The equations of motion of the resulting non-relativistic field theory are given by the curved space generalization of the so-called L\'evy-Leblond equations \cite{LevyLeblond:1967zz}
\begin{align}\begin{split}
  \gamma^0\big(\tilde D_0 +{\rm{i}}\,q\,\tau^\mu A_\mu\big)\psi_+ +\gamma^a\big(\tilde D_a +{\rm{i}}\,q\,e^\mu{}_aA_\mu\big)\psi_- &=0\,, \\[.1truecm]
  \gamma^a\big(\tilde D_a +{\rm{i}}\,q\,e^\mu{}_a A_\mu\big)\psi_+ -2\,m\,\psi_- &=0\,.
\end{split}\end{align}
The second equation can be used to solve for the auxiliary spinor $\psi_-$ and eliminate it from the Lagrangian \eqref{4dNRdirac}. Substituting the solution for $\psi_-$
back into the first equation
we obtain the curved space generalization of the Schr\"odinger-Pauli equation, including an electromagnetic coupling:
\begin{align} \label{modLL}
 \Big[\gamma^0\big(\tilde D_0 +{\rm{i}}\,q\,A_0\big) +\frac1{2m}\,\big(\tilde  D_a +{\rm{i}}\,q\,A_a\big)^2
        +\frac{{\rm{i}}\,q}{2m}\,\gamma^{ab}\,\big(\tilde D_aA_b\big)\Big]\psi_+ =0 \,.
\end{align}
The $\tilde D_a$ in the last term of this equation acts only on $A_b$ and in flat space the Land\'e factor of the electron can be read off from this term.

\subsection{Spin-1}

In order to obtain non-relativistic field equations for a massive
spin-1 field in an arbitrary background, we start from a relativistic
theory for a complex Proca field $A_\mu$,
minimally coupled to gravity and the background gauge field $M_\mu$
\begin{align}
 E^{-1} \mathcal{L}_{\rm rel} = -\frac14\,g^{\mu\rho} g^{\nu\sigma}F_{\mu\nu}^*F_{\rho\sigma} -\frac12\,M^2\, g^{\mu\nu} A_\mu^* A_\nu  \,,
\end{align}
with
\begin{align}
 F_{\mu\nu} = 2\,D_{[\mu}A_{\nu]} = 2\,\partial_{[\mu}A_{\nu]} - 2\, {\rm{i}}\,M\,M_{[\mu}A_{\nu]}\,.
\end{align}
Apart from diffeomorphism invariance, this Lagrangian is also invariant under a U(1) symmetry
\begin{equation}
\delta A_\mu = {\rm{i}} \, M \, \Lambda \, A_\mu \,,
\end{equation}
whose physical interpretation is as before.
Applying the redefinitions of section \ref{sec:bgcontr} and rescaling $M = \omega m$, one finds again that the limit $\omega \rightarrow \infty$ is well-defined. The resulting non-relativistic Lagrangian is given by
\begin{align} \label{eq:massvecnonrel1}
 e^{-1} \mathcal{L}_{\rm non-rel} = -\frac14\,\tilde{F}_{ab}^*\tilde{F}^{ab} -\frac12\, {\rm{i}} \,m\,A^*_a \tilde{F}_{a0} +\frac12\,{\rm{i}}\,m\,A_a\,\tilde{F}_{a0}^* +\frac12\,m^2\,\big|A_0\big|^2 \,,
\end{align}
where now
\begin{align} \label{eq:massvecnonrel2}
 \tilde{F}_{\mu\nu} =  2\, \tilde{D}_{[\mu} A_{\nu]} = 2\, \partial_{[\mu} A_{\nu]} +2 {\rm{i}}\,m\,m_{[\mu}\,A_{\nu]} \,.
\end{align}
As before, this Lagrangian is invariant under diffeomorphisms as well as under a U(1) central charge transformation given by
\begin{equation}
  \delta A_\mu = {\rm{i}} \, m \, \sigma \, A_\mu \,.
\end{equation}
The physical content of this Lagrangian is most easily seen by restricting to a flat background and examining the resulting equations of motion. These are given by
\begin{align}\begin{split}
 {\rm{i}}\, \partial^iA_i -m\,A_0 &= 0 \,, \\
 {\rm{i}} \,m\,\partial_iA_0 -2\, {\rm{i}} \,m\,\partial_t A_i -\partial^jF_{ji} &= 0 \,.
\end{split}\end{align}
Applying $\partial^i$ on the second equation of these and using the first equation to rewrite the ensuing $\partial_t \partial^i A_i$ term as a $\partial_t A_0$ term, one finds a Schr\"odinger equation for $A_0$. Similarly, the last term of the second equation contains a $\partial_i \partial^j A_j$ term that can be rewritten as a $\partial_i A_0$ term using the first equation. One then finds that the second equation reduces to three Schr\"odinger equations for each component of $A_i$. In the flat background, the diffeomorphisms and U(1) central charge symmetry of the non-relativistic theory reduce to the following transformation rules:
\begin{align}\begin{split}
 \delta A_0 &= \Big(\zeta\partial_t +\xi^i\partial_i -t\,\lambda^i\partial_i -x^j\lambda^i{}_j\,\partial_i
           +{\rm{i}}\,m\,\lambda^ix^i -{\rm{i}}\,m\,\sigma \Big)A_0  -\lambda^iA_i \,, \\
 \delta A_i &= \Big(\zeta\partial_t +\xi^j\partial_j -t\,\lambda^j\partial_j -x^k\lambda^j{}_k\,\partial_j
           +{\rm{i}}\,m\,\lambda^jx^j  -{\rm{i}}\,m\,\sigma\Big)A_i  +\lambda_i{}^jA_j \,.
\end{split}\end{align}
Note that, apart from the usual transformations expected of Schr\"odinger fields, there is also a non-trivial boost under which $A_0$ transforms to $A_i$. Furthermore $A_i$ transforms as a vector under spatial rotations.

\subsection{Spin-3/2}

Finally, let us consider the action of a massive spin-3/2 field in four dimensions, minimally coupled to $M_\mu$ and a gravitational background:
\begin{align}
 E^{-1}\mathcal{L}_{\rm rel} = \bar\Psi_\mu \,\gamma^{\mu\nu\rho}\,D_\nu\Psi_\rho -M\,\bar\Psi_\mu\,\gamma^{\mu\nu}\Psi_\nu +{\rm h.c.}
\end{align}
We have defined the covariant derivative of a vector-spinor as
\begin{align}\label{PsiMcouple}
 D_\mu\Psi_\nu = \partial_\mu\Psi_\nu -\frac14\,\Omega_\mu{}^{AB}\gamma_{AB}\Psi_\nu -{\rm i}\,M\,M_\mu\,\Psi_\nu \,,
\end{align}
in line with its transformation properties under local Lorentz and U(1) transformations:
\begin{align}
 \delta \Psi_\mu = \frac14\,\Lambda^{AB}\gamma_{AB}\Psi_\mu +{\rm i}\,M\,\Lambda\,\Psi_\mu \,.
\end{align}
Note that \eqref{PsiMcouple} entails the coupling of the relativistic
spinor to the gauge field $M_\mu$
in the same way as shown in all previous cases. The non-relativistic
result is then obtained by using the redefinitions of the background
fields given in section \ref{sec:bgcontr}, the same projections of
$\Psi_\mu$
onto spinors $\Psi_{\mu \pm}$ as in (\ref{eq:spinproj}) and the same $\omega$-dependent redefinition to spinors $\psi_{\mu \pm}$ as in (\ref{eq:spinprojresc}).
After sending $\omega\to\infty$
we obtain the following non-relativistic action for $\psi_{\mu \pm}$:
\begin{align}\begin{split}\label{NR_RS-field}
 e^{-1}\mathcal{L}_{\rm NR} &= e^\mu{}_ae^\nu{}_be^\rho{}_c\,\big(\bar\psi_{\mu+}\gamma^{abc}\tilde D_\nu\psi_{\rho-}
                                                          +\bar\psi_{\mu-}\gamma^{abc}\tilde D_\nu\psi_{\rho+}\big)
         +3\,\tau^{[\mu} e^\nu{}_ae^{\rho]}{}_b\,\bar\psi_{\mu+}\gamma^{ab0}\tilde D_\nu\psi_{\rho+} \\
  &\quad -2m\,\big(e^\mu{}_ae^\nu{}_b\,\bar\psi_{\mu-}\gamma^{ab}\psi_{\nu-}
                  -2\,\tau^{[\mu}e^{\nu]}{}_a\,\bar\psi_{\mu-}\gamma^{a0}\psi_{\nu+} \big) +{\rm h.c.}\,,
\end{split}\end{align}
where we have used
\begin{align}\begin{split}
 \tilde D_\mu\psi_{\nu+} &= \partial_\mu\psi_{\nu+} -\frac14\,\omega_\mu{}^{ab}\gamma_{ab}\psi_{\nu+}
           +{\rm i}\,m\,m_\mu\,\psi_{\nu+} \,, \\
 \tilde D_\mu\psi_{\nu-} &= \partial_\mu\psi_{\nu-} -\frac14\,\omega_\mu{}^{ab}\gamma_{ab}\psi_{\nu-}
           +\frac12\,\omega_\mu{}^a\gamma_{a0}\psi_{\nu+} +{\rm i}\,m\,m_\mu\,\psi_{\nu-} \,.
\end{split}\end{align}
One can check by explicit calculation that the Lagrangian \eqref{NR_RS-field} is invariant under Galilean boosts and other
symmetries are manifest. The transformation rules of the non-relativistic spinors are given by
\begin{align}\begin{split}\label{delta32}
 \delta \psi_{\mu+} &= \frac14\,\lambda^{ab}\gamma_{ab}\psi_{\mu+} -{\rm i}\,m\,\sigma\,\psi_{\mu+} \,, \\
 \delta \psi_{\mu-} &= \frac14\,\lambda^{ab}\gamma_{ab}\psi_{\mu-} -\frac12\,\lambda^a\gamma_{a0}\psi_{\mu+}
            -{\rm i}\,m\,\sigma\,\psi_{\mu-} \,.
\end{split}\end{align}
This concludes our discussion of the non-relativistic limit of massive fields.

\section{Massless fields} \label{sec:massless}

In this section, we will discuss how the background contraction,
extended to include dynamical fields, also leads to Lagrangians for
massless fields in arbitrary Newton-Cartan backgrounds without
torsion. As mentioned in \cite{Leblond:1963},
massless representations that correspond to true (instead of ray)
representations of the Galilei group exist. Viewed as representations
of the Bargmann algebra, they thus do not transform non-trivially
under the central charge. Massless non-relativistic fields can
therefore be real. Furthermore, when coupled to an arbitrary
background, no non-trivial couplings to the central charge gauge field
$m_\mu$
are expected to appear. Consequently, one can obtain such field
theories starting from relativistic Lagrangians for real fields, that
cannot be coupled to $M_\mu$
and that thus only involve couplings to the gravitational
background. In this respect it is useful to recall that in the massive
case, the couplings to $M_\mu$
were necessary to cancel the diverging mass terms in the
$\omega \to \infty$
limit. In the absence of any mass term, the $\omega \to \infty$
is thus well-defined without the need for additional minimal couplings
to $M_\mu$.

Here, we will extend the background contraction to Lagrangians that
describe single real massless fields of spins 0 up to 3/2. This leads
to simple non-relativistic field theories, where the fields transform
trivially under Galilean boosts. It is interesting to consider
slightly more involved cases, where non-trivial boost behavior for
the fields is retained. This is in particular possible in the spin-1/2, spin-1 and spin-3/2 cases and we will comment on this in the following.

\subsection{Spin-0}

Starting from the Lagrangian for a real, massless, relativistic scalar field
\begin{equation}
  \label{eq:mlscalrel}
  E^{-1} \mathcal{L}_{\rm rel} = - \frac12\, g^{\mu \nu} \partial_\mu \Phi\, \partial_\nu \Phi \,,
\end{equation}
and applying the redefinitions of section \ref{sec:bgcontr}, one finds that the $\omega \rightarrow \infty$ limit is well-defined and leads to the following non-relativistic Lagrangian
\begin{equation}
  \label{eq:mlscalnrel}
  e^{-1} \mathcal{L}_{\rm non-rel} = -\frac12\, \partial_a \Phi\, \partial^a \Phi \,.
\end{equation}
One is thus led to a field obeying a simple Poisson equation, which is the unique Galilean covariant field equation one can write down for a real scalar field involving up to two derivatives.

Note that one could also take the $m \rightarrow 0$ limit in the Lagrangian (\ref{nrscalar}) for a massive Schr\"odinger field. This limit leads to a Poisson equation for a complex field, or equivalently for two real fields. Another way of getting the result (\ref{eq:mlscalnrel}) from (\ref{nrscalar}) is obtained by noting that one can gauge fix the central charge transformation by requiring $\Phi$ to be real. Indeed, writing the complex field $\Phi$ in a polar decomposition
\begin{equation}
  \Phi = {\rm{e}}^{{\rm{i}} \, \phi} |\Phi| \,,
\end{equation}
one sees that the central charge symmetry acts as a St\"uckelberg shift symmetry on $\phi$ and can therefore be fixed by putting $\phi = 0$. Doing this in (\ref{nrscalar}) then leads to the above Lagrangian for a massless real field.

\subsection{Spin-1/2}

The simplest possibility to obtain a non-relativistic spin-1/2 field coupled to an arbitrary background is to start from a real (i.e.~Majorana), massless, relativistic spin-1/2 field with Lagrangian
\begin{equation}
  \label{eq:majspin12rel}
  E^{-1} \mathcal{L}_{\rm rel} = \bar{\Psi} \slashed{D} \Psi\,,
\end{equation}
where the covariant derivative contains the spin connection of the relativistic background
\begin{equation}
  D_\mu \Psi = \partial_\mu \Psi - \frac14\, \Omega_\mu{}^{AB} \gamma_{AB} \Psi \,.
\end{equation}
This Lagrangian has a well-defined $\omega \rightarrow \infty$ limit, given by
\begin{equation}
  e^{-1} \mathcal{L}_{\rm non-rel} = \bar{\Psi}\, \gamma^a \tilde{D}_a \Psi \,,
\end{equation}
where the covariant derivative contains the Newton-Cartan spin connection of spatial rotations
\begin{equation}
  \tilde{D}_\mu \Psi = \partial_\mu \Psi - \frac14\, \omega_\mu{}^{ab} \gamma_{ab} \Psi \,,
\end{equation}
since the non-relativistic $\Psi$ field only transforms under spatial rotations and no longer under boosts.

It is also possible to obtain a Lagrangian for a non-relativistic massless spin-1/2 field, that transforms non-trivially under Galilean boosts by simply taking the $m\rightarrow 0$ limit of eqs.~(\ref{4dNRdirac})--(\ref{modLL}). Note that in this case, one works with a Dirac spinor. The relativistic starting point is thus U(1) invariant, but we do not gauge this U(1) invariance using the background gauge field $M_\mu$. Introducing $M_\mu$-couplings would lead to terms that diverge in the $\omega \rightarrow \infty$ limit. Working with a Dirac spinor however has the advantage that one can introduce non-trivial boost transformations via the projections (\ref{eq:spinproj}) and rescalings (\ref{eq:spinprojresc}).

\subsection{Spin-1}

The background contraction procedure allows one to easily obtain a non-relativistic theory for a vector field coupled to an arbitrary non-relativistic background, starting from the Lagrangian of a real, massless, relativistic vector field
\begin{equation}
  \label{eq:realvecrel}
  E^{-1} \mathcal{L}_{\rm rel} = - \frac14\, g^{\mu\rho} g^{\nu\sigma} F_{\mu\nu} F_{\rho \sigma} \,,
\end{equation}
where $F_{\mu\nu}$ is the usual Maxwell field strength.
Simply contracting the background according to the redefinitions of section \ref{sec:bgcontr} and not rescaling the vector field itself, leads to the following non-relativistic Lagrangian
\begin{equation}
  \label{eq:realvecnonrel}
  e^{-1} \mathcal{L}_{\rm non-rel} = - \frac14\, F_{ab} F^{ab} \,.
\end{equation}
Note that this Lagrangian only contains the spatial part $A_a$ of the vector field $A_\mu$ and not the electric potential $A_0$. By taking the $m \rightarrow 0$ limit of (\ref{eq:massvecnonrel1}) and (\ref{eq:massvecnonrel2}), one obtains a Lagrangian of the form of (\ref{eq:realvecnonrel}) for a complex vector field, or equivalently for two real vector fields.

One can, however, obtain a non-relativistic Lagrangian that contains the electric potential $A_0$ as well,
by starting from a relativistic Maxwell field and a massless scalar
\footnote{The motivation for this particular combination stems from considerations in supersymmetric theories, i.e.~non-relativistic
limits of $\mathcal{N}=2$ vector supermultiplets, see e.g.~\cite{Bergshoeff:2015ija}, and their corresponding actions.}
\begin{equation}
  \label{eq:realvecscalrel}
E^{-1} \mathcal{L}_{\rm rel} = - \frac14\, g^{\mu\rho} g^{\nu\sigma} F_{\mu\nu} F_{\rho \sigma} - \frac12\, g^{\mu\nu} \partial_\mu \rho\, \partial_\nu \rho \,.
\end{equation}
Introducing redefined fields as
\begin{equation} \label{eq:AB}
  A = E^\mu{}_0 A_\mu - \rho \,, \qquad B = E^\mu{}_0 A_\mu + \rho \,,
\end{equation}
one can apply the background redefinitions of section \ref{sec:bgcontr}, supplemented with the rescalings
\begin{equation} \label{eq:ABtilde}
  A = \frac{1}{\omega}\, \tilde{A} \,, \qquad B = \omega\, \tilde{B} \,,
\end{equation}
to obtain the following non-relativistic Lagrangian in the $\omega \rightarrow \infty$ limit
\begin{equation}
  \label{eq:realvecscalnonrel}
  e^{-1} \mathcal{L}_{\rm non-rel} = \frac18\,\partial_0\tilde{B}\,\partial_0\tilde{B}
              +\frac12\,\tilde{D}_a\tilde{A}\,\partial^a\tilde{B} -\frac14\,F_{ab} F^{ab}
              -\frac12\,\tilde{D}^a A_a\, \partial_0 \tilde{B} \,,
\end{equation}
where the following derivatives were used
\begin{align} \label{eq:halfcovders}
  \tilde{D}_\mu \tilde{A} &= \partial_\mu \tilde{A} + \omega_\mu{}^a A_a \,, \nonumber \\
  \tilde{D}_\mu A_a &= \partial_\mu A_a - \omega_{\mu\, a}{}^b A_b + \frac12\, \omega_\mu{}^a \tilde{B} \,.
\end{align}
Note that contrary to the massive spin-1 case, here we do not consider a vector field $A_\mu$ with curved space-time indices as the fundamental variables. Rather, we work with a spatial vector $A_a$ with spatial flat indices and two extra fields $\tilde{A}$, $\tilde{B}$. As a consequence, these fields transform non-trivially under local spatial rotations and Galilean boosts, while they transform as scalars under general coordinate transformations. The transformation rules under local spatial rotations and Galilean boosts can be derived using e.g.~(\ref{eq:AB}) and (\ref{eq:ABtilde}) following the procedure of section \ref{sec:bgcontr} and \cite{Bergshoeff:2015uaa} and are given by
\begin{align}
  \delta \tilde{A} &= -\lambda^a A_a \,, \qquad \qquad \qquad \delta \tilde{B} = 0 \nonumber \\
  \delta A_a &= \lambda_a{}^b A_b - \frac12\, \lambda^a \tilde{B} \,.
\end{align}
It is with respect to these transformations that the above derivatives (\ref{eq:halfcovders}) are defined. The above Lagrangian is also invariant under the U(1) gauge transformation
\begin{equation}
  \label{eq:nonrelU1}
  \delta \tilde{A} = \tau^\mu \partial_\mu \Lambda \,, \qquad\quad \delta A_a = e^\mu{}_a \partial_\mu \Lambda \,.
\end{equation}
This invariance is not manifest for the Lagrangian (\ref{eq:realvecscalnonrel}). By twice partially integrating the last term of this Lagrangian, one can however easily rewrite it in terms of the gauge invariant quantities
\begin{equation}
  \tilde{D}_{[a} A_{b]} \,, \qquad \qquad \tilde{D}_a \tilde{A} - \tilde{D}_0 A_a \,.
\end{equation}
Note that in performing these two partial integrations, one has to commute $\tilde{D}_0$ and $\tilde{D}_a$ derivatives. The commutator contributions $[\tilde{D}_0, \tilde{D}_a] A^a$ that one thus picks up are however vanishing upon using the background equations of motion.

To get a better physical understanding of the Lagrangian (\ref{eq:realvecscalnonrel}), we consider the equations of motion when restricted to a flat background
\begin{align}\begin{split}
  \partial^i \partial_i \tilde{B} &= 0 \,, \\
  \partial_i \partial_t \tilde{B} + \partial^j F_{ji} &= 0 \,, \\
  \partial_t \partial_t \tilde{B} - 2\, \partial^i \partial_i \tilde{A} + 2\, \partial_t \partial^i A_i &= 0 \,.
\end{split}\end{align}
One can consistently put $\tilde{B}$ to zero, as it amounts to a scalar field that is inert under gauge transformations. The remaining equations for $\tilde{A}$ and $A_i$ then coincide with the equations of Galilean Electromagnetism in the magnetic limit \cite{leBellac,Bagchi:2014ysa}, where $\tilde{A}$ plays the role of the electric potential. As was shown in \cite{Duval:2009vt,Bagchi:2014ysa}, this theory is not only invariant under the Galilei group, but also under the Galilean conformal group. The latter is the conformal extension of the Galilei group that is obtained by performing an In\"on\"u-Wigner contraction of the relativistic conformal group. Since the relativistic Lagrangian we started from is conformally invariant when restricted to flat space, it is not surprising to see that the non-relativistic limit is invariant under Galilean conformal symmetry.

\subsection{Spin-3/2}

Like in the spin-1/2 case, the simplest way to obtain a Lagrangian for a massless vector-spinor is to start with the Lagrangian of a relativistic, massless Majorana vector-spinor,
\begin{equation}
 E^{-1}\mathcal{L_{\rm rel}}=\bar\Psi_\mu\gamma^{\mu\nu\rho}D_\nu\Psi_\rho \,,
\end{equation}
where the (anti-symmetrized) covariant derivative is given by
\begin{equation}
D_{[\mu}\Psi_{\nu]}= \partial_{[\mu}\Psi_{\nu]} -\frac14\,\Omega_{[\mu}{}^{AB}\gamma_{AB}\Psi_{\nu]} \,.
\end{equation}
The $\omega\rightarrow\infty$ limit of this Lagrangian is well-defined and leads to
\begin{align}
e^{-1}\mathcal{L}_{\rm non-rel}= e^\mu{}_ae^\nu{}_be^\rho{}_c\,\bar\Psi_\mu\gamma^{abc}\tilde D_\nu\Psi_\rho \,,
\end{align}
where the (anti-symmetrized) covariant derivative of the non-relativistic vector-spinor is
\begin{align}
 \tilde D_{[\mu}\Psi_{\nu]} = \partial_{[\mu}\Psi_{\nu]} -\frac14\,\omega_{[\mu}{}^{ab}\gamma_{ab}\Psi_{\nu]} \,.
\end{align}
Another formulation with a vector-spinor that transforms non-trivially under boosts can be obtained by taking the $m\to0$ limit of eqs.~(\ref{NR_RS-field})--(\ref{delta32}). Like in the massive case we would then work with Dirac spinors that have an extra global U(1) symmetry which, however, we do not gauge by coupling to $M_\mu$.

\section{Summary and Outlook} \label{sec:concl}

In this paper, we have shown how non-relativistic Lagrangians for free massive and massless dynamical fields of spins 0 up to 3/2 coupled to arbitrary torsionless Newton-Cartan backgrounds can be obtained from their relativistic counterparts. We have done so by extending the procedure developed in \cite{Bergshoeff:2015uaa}, by which non-relativistic Newton-Cartan backgrounds are derived from arbitrary relativistic ones via an In\"on\"u-Wigner-like contraction, to include dynamical fields. The relativistic backgrounds of \cite{Bergshoeff:2015uaa} include a curl-free U(1) gauge field $M_\mu$, needed to make the contraction procedure well-defined. In the present paper, this field acquires extra significance when considering massive fields, that are necessarily complex. In those cases, the relativistic theory one starts from includes not only gravitational minimal couplings but also minimal couplings to $M_\mu$ that gauge the U(1) symmetry associated to the conservation of the number of particles minus the number of antiparticles. Starting from this, the contraction yields the correct couplings to the central charge gauge field $m_\mu$ of the Newton-Cartan background, whose role is to gauge the symmetry associated to particle number conservation.

Although related results for massive spin-0 and spin-1/2 fields have already appeared in the literature \cite{Jensen:2014wha,Fuini:2015yva,Geracie:2015dea}, here we have re-derived these by extending the method of \cite{Bergshoeff:2015uaa} and we have shown that in this way similar results can be obtained for massive fields of all spins up to 3/2, in a uniform and straightforward manner. Even though we have mostly restricted ourselves to free fields, it is in principle rather straightforward to include interactions (as we have for instance done in the massive spin-1/2 case). Our procedure can moreover also be straightforwardly applied to massless fields. Starting from relativistic Lagrangians for single massless real fields of a certain spin, one typically ends up with fields that obey equations that can be viewed as curved space generalizations of the Poisson equation. By adopting a less restrictive starting point that includes multiple real fields, it is possible to obtain more interesting massless non-relativistic theories, such as Galilean Electromagnetism in the magnetic limit \cite{leBellac} in the spin-1 case.

Finally, let us mention some interesting generalizations and directions for future research. As we have mentioned in the massless
spin-1 case, our contraction procedure leads to massless field theories that exhibit Galilean conformal invariance, upon
restriction to flat space. In view of the significance of the Galilean conformal group in attempts to formulate a holographic
correspondence in asymptotically flat space-times \cite{Bagchi:2010eg,Bagchi:2010zz}, it would be interesting to see whether
the method described here can be used to obtain non-trivial, interacting Galilean conformal field theories. In this paper, we
have restricted ourselves to field theories coupled to non-dynamical Newton-Cartan backgrounds. It would be interesting to see
how dynamical Newton-Cartan gravity can be included. Since the results described here hold only for minimally coupled field
theories, it could be of interest to also explore whether non-minimal couplings of fields to Newton-Cartan backgrounds can be
obtained in a similar manner.
Our limiting procedure gives rise to non-relativistic field theories coupled to torsionless backgrounds. The recipe to couple to
torsionfull backgrounds instead would be in a first step to generalize the Newton-Cartan background fields here to those of the
Schr\"odinger theory, see \cite{Bergshoeff:2014uea}, and gauge fixing dilations afterwards. This leads to non-relativistic
field theories coupled to twistless torsional Newton-Cartan backgrounds.
Finally, in view of recent developments in supergravity versions of Newton-Cartan gravity
\cite{Andringa:2013mma,Bergshoeff:2015uaa,Bergshoeff:2015ija}, it is interesting to generalize the results of this paper to
non-relativistic supersymmetric theories coupled to Newton-Cartan supergravity backgrounds. These results could then be useful
for discussing matter coupled Newton-Cartan supergravity theories, see \cite{Bleeken:2015ykr}, and supersymmetric localization for non-relativistic theories,
see also \cite{Knodel:2015byb}. We hope to report on these issues in the future.

\section*{Acknowledgements}

It is a pleasure to thank Dieter Van den Bleeken for useful discussions. We also want to thank Quim Gomis for clarifying discussions in the 
early stages of this project and for useful comments on a first version of this paper. JR was supported by the NCCR SwissMAP, funded by the Swiss National Science Foundation. TZ acknowledges financial support by the Dutch Academy of Sciences (KNAW).

\small{


\providecommand{\href}[2]{#2}\begingroup\raggedright\endgroup

}

\end{document}